\documentclass[aps,prl,showpacs,preprint,amssymb,amsmath,floatfix]{revtex4}
\usepackage{graphicx,bm}
\begin{document}
\bibliographystyle{prsty}
\title{Lifetimes of Stark-shifted image states}
\author{S. Crampin}\email{s.crampin@bath.ac.uk}
\affiliation{Department of Physics, University of Bath, Bath BA2 7AY,
United Kingdom}
\date{\today}
\begin{abstract}

The inelastic lifetimes of electrons in image-potential states at Cu(100)
that are 
Stark-shifted by the electrostatic tip-sample interaction in the scanning
tunneling microscope are calculated using the many-body GW approximation.
The results demonstrate that in typical tunneling conditions the image state
lifetimes are significantly reduced from their field-free values. 
The Stark-shift to higher energies increases the number of inelastic scattering
channels that are available for decay, with field-induced changes in 
the image state
wave function increasing the efficiency of the inelastic scattering through
greater overlap with final state wave functions.

\end{abstract}

\pacs{73.20.At,68.37.Ef,72.15.Lh}

\maketitle

\section{Introduction}
\label{sec:intro}

The scanning tunneling microscope (STM) is a versatile and powerful probe of
surface electronic structure; but it is not ideal. The electric field between 
the probe tip and the surface of the sample affects the surface. 
This influence can be exploited to positive effect, most dramatically
through the controlled modification of surface atomic 
structure \cite{eig90_,cro93_}.
More prosaically, the influence of the tip must be allowed for when
interpreting STM measurements, especially at semiconductor surfaces where
tip-induced band bending occurs \cite{mce93_}.

Recently a significant Stark-effect -- the shift in energy due to the electric 
\begin{figure*}[b!]
\includegraphics[bbllx=63,bblly=334,bburx=519,bbury=502,
width=150mm,clip=]{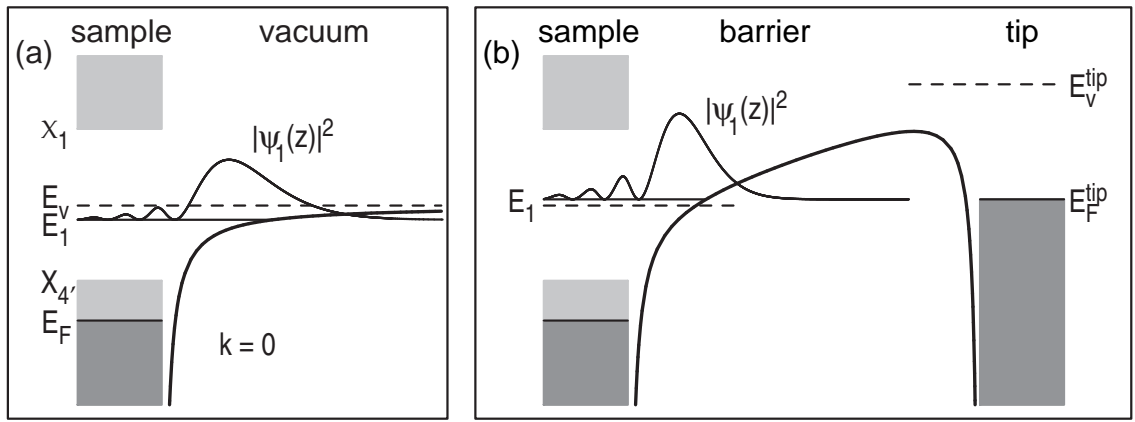}
\caption{
\label{fig:energy}
(a) Schematic energy diagram for the potential at the Cu(100) surface. 
The vacuum level lies near the center of the projected band gap which at
${\bm k}_\|=0$ lies between the energies of the bulk $X_{4'}$ and $X_1$ levels.
The potential due to the image interaction (thick solid line) 
produces a hydrogenic-like
series of excited states converging on the vacuum energy (dashed line); 
the $n=1$ level
and corresponding probability density are shown. (b) 
Energy diagram for the surface in the presence of the tip of the
scanning tunneling microscope, at the threshold bias for tunneling into
the $n=1$ image state. The field in the tunnel junction
causes a Stark shift of the image state spectrum, 
with concomitant modification of the probability densities of the states.}
\end{figure*}
field -- has been identified in scanning tunneling spectroscopy (STS) of
Shockley surface-state electrons at a metal surface \cite{lim03_}. 
Surface states, in which 
the electron is caught between the barrier potential outside the
surface and a band gap in the crystal, have been extensively studied by
STM and STS, with a particular focus on their 
dynamics \cite{jli98_,kli00a,bur99_,bra02_} and 
interactions \cite{rep00_}, 
and the recognition of a Stark-shift in the case of Shockley
states reconciles a discrepancy that has existed
between STS-derived binding energies and those from photoelectron 
spectroscopy \cite{rei01_}. A more pronounced Stark-effect has been 
known for some time \cite{bec85_,bin85_} in the case of a second 
class of surface electron state, namely the image-potential states that arise 
when an electron outside a conductor
polarises the surface and is attracted to the resulting ``image charge'', shown 
schematically in Fig. \ref{fig:energy}.
Image-potential states are more weakly bound than Shockley surface states (which
lie close to the Fermi energy $E_F$),
forming a hydrogenic-like series with energies
\begin{equation}
E_n=E_{\textrm{v}}-\frac{0.85\ \textrm{eV}}{(n+a)^2}, \quad n=1,2,\dots
\label{eqn:En}
\end{equation}
converging on the vacuum level of the surface $E_{\textrm{v}}$. 
In (\ref{eqn:En}) $a$ is a quantum 
defect that depends upon the surface. 
Tunneling via image states requires
significantly greater bias voltages than Shockley states, and the image
state electrons are Stark-shifted to higher energies by several tenths
of an eV \cite{bec85_,bin85_,wah03_}.

The presence of a measurable Stark-shift in the surface state energies
raises the important question as to whether there are also changes in the
inelastic interactions of the surface state electrons in the presence of
the STM tip. Electronic excitations in the surface state bands decay on a
femtosecond timescale through interactions with the electrons and phonons
of the surface and bulk, and there has been considerable activity in
recent years directed at an understanding of these
interactions \cite{ech04_}. A significant electric-field induced lifetime change
would have important consequences for the interpretation of STM and STS
experiments investigating the dynamical properties of image-potential
states, for example in nanostructures where the lateral resolution of the 
the STM is paramount.

To investigate this issue we have performed many-body calculations
of the lifetimes of image-potential states at Cu(100) in the presence 
of an electric-field due to the tip of an STM.
Our calculations are based upon the approach introduced by 
Chulkov {\it et al}. \cite{chu98_} and used subsequently in numerous 
surface state lifetime studies  with considerable success \cite{ech04_}. 
The damping rate or inverse
lifetime of the image state is calculated from the expectation value
of the imaginary part of the non-local self energy operator
\begin{equation}
\Gamma=\hbar/\tau=-2\int\!d{\bm r}\!\int d{\bm r}'
\psi^\ast({\bm r}) \textrm{Im} \Sigma({\bm r},{\bm r}';E)
\psi({\bm r}').
\label{eqn:gamma}
\end{equation}
The energy of the image state is $E$, and $\psi$ the wave function.
The imaginary part of the self energy is calculated in the GW approximation
of many-body theory, which uses the first term only in the series expansion of
$\Sigma$ in terms of the screened Coulomb interaction $W$: 
\begin{equation}
\textrm{Im} \Sigma({\bm r},{\bm r}';\epsilon)=
-\frac{1}{\pi}\int_{E_F}^{\textstyle \epsilon} \!\!d\epsilon'
\textrm{Im} G({\bm r},{\bm r}';\epsilon')
\textrm{Im} W({\bm r},{\bm r}';\epsilon\!-\!\epsilon').
\label{eqn:sigma}
\end{equation}
We use the zero'th order approximation to the Green function; in the
spectral representation
\begin{equation}
G({\bm r},{\bm r}';\epsilon)=\sum_i\frac
{\psi_i({\bm r})\psi_i^\ast({\bm r}')}{\epsilon-E_i+i\delta}
\end{equation}
where the $\psi_i({\bm r})$ are one-electron eigenfunctions with
eigenenergies $E_i$ and $\delta$ a positive infinitesimal.
The screened interaction is evaluated in the random phase approximation
(RPA)
\begin{eqnarray}
W({\bm r},{\bm r}';\omega)&=&V({\bm r}\!-\!{\bm r}')+
\int d{\bm r}_1\!\int d{\bm r}_2 V({\bm r}\!-\!{\bm r}_1)
\nonumber\\
&&\times
\chi^0({\bm r}_1,{\bm r}_2;\omega)W({\bm r}_2,{\bm r}';\omega),
\label{eqn:W}
\end{eqnarray}
where $V({\bm r})$ is the bare Coulomb interaction and
$\chi^0({\bm r},{\bm r}';\omega)$ is the density-density response function of
the non-interacting electron system:
\begin{eqnarray}
\chi^0({\bf r},{\bf r}';w)&=&
-\frac{2}{\pi}\int^{E_F}\!\!d\epsilon
\ \textrm{Im}G({\bf r},{\bf r}';\epsilon)\nonumber\\
&&\times\left[
G({\bf r},{\bf r}';\epsilon\!+\!\omega)+
G^\ast({\bf r},{\bf r}';\epsilon\!-\!\omega)\right].
\label{eqn:chi0}
\end{eqnarray}
This GW-RPA approach has been shown to give decay rates 
for image states at Cu surfaces that are within 1 meV
of those found using the more complete 
GW$\Gamma$-TDLDA approximation \cite{sar99_},
in which exchange-correlation effects that are omitted in GW-RPA 
are included in both (\ref{eqn:W}) and (\ref{eqn:chi0}).
Note that the phonon contribution to the decay rate 
of image-potential states at Cu(100) is $< 0.5$ meV and so
safely ignored here \cite{eig03_}.

We calculate the Green function using a one-dimensional
pseudopotential which by construction reproduces
the Cu(100) bulk band edge energies $X_{4'}$ and $X_1$, and the 
energies of the unoccupied surface resonance and first image state
at the field-free surface, and which also
accurately predicts the energies of the
higher image states \cite{chu99_}. 
In the direction parallel to the surface we
assume parabolic dispersion with effective masses ($m^\ast$) fitted to 
ab-initio
band structures. To model the influence of the STM we follow Limot 
{\it et al.} \cite{lim03_} and include a linear
potential due to the bias voltage between the STM tip and the sample,
and modify the image potential to include the multiple 
images present in the tunnel junction geometry. Using this potential 
we are able to 
reproduce the sequence of Stark-shifted image state energies and 
increments in tip-sample distance observed at Cu(100) in $z(V)$ spectroscopy by
Wahl \textit{et al}., \cite{wah03_}  who were also able to describe
them using a model that omitted the effect of multiple-images.

\begin{figure}[t!]
\includegraphics[width=75mm,clip=]{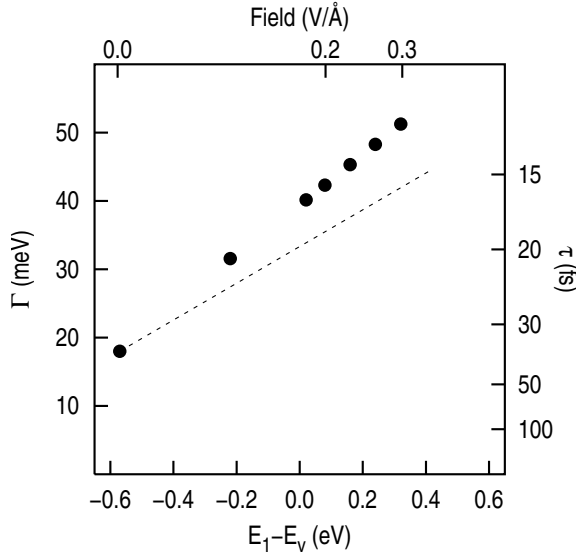}
\caption{
\label{fig:gamma}
Calculated decay rates ($\Gamma$) and lifetimes ($\tau$)
of the Stark-shifted $n=1$
image state at Cu(100) (filled circles). 
$E_1$ is the image state energy in the presence 
of the electric field, and $E_{\textrm{v}}$ 
the field-free vacuum level of Cu(100).
The significance of the dashed line is explained in the text.
}
\end{figure}

In Fig. \ref{fig:gamma} we show the calculated damping rates (inverse lifetimes)
of the Stark-shifted $n=1$ image potential state at Cu(100). 
Calculation parameters have been systematically varied to ensure that
decay rates are converged to within 1 meV.
At the field-free Cu(100) surface the image state lies 
at an energy of $-0.57$ eV
relative to the vacuum energy, and we find $\Gamma=18\pm 1$\ meV 
for this state,
corresponding to 
a lifetime of $\tau=37\pm 2$\ fs. This compares well with the lifetime
$\tau=40\pm 6$\ fs measured using
time-resolved two-photon photoemission (2PPE)
\cite{hof97_,ber02_},
and $\tau=38$\  fs found in previous GW calculations \cite{sar99_}. 
In the presence of the electric field due to the STM tip, 
the image state electrons
are Stark-shifted to higher energies. At currents of 0.1-1\ nA 
the $n=1$ level is 
observed in $z(V)$ spectroscopy at bias voltages near $4.7$ V, 
corresponding to an energy $E_1 \simeq E_{\textrm{v}}+0.1$ eV
\cite{wah03_}. For the results in Fig. \ref{fig:gamma}
the tip-sample separation has been varied so that
the bias voltage coincides with the $n=1$ image state energy in the 
presence of the electric field. This situation corresponds to
the onset of tunneling into the $n=1$ level (Fig. \ref{fig:energy}(b)). 
In these conditions the resulting image state energy increases linearly 
with the applied electric field, and we find that the decay 
rates also increase linearly. The rate of change is 
$d\Gamma/dE_1=0.037\pm 0.002$, so that when the image state
is Stark-shifted to $E_1=E_{\textrm{v}}+0.1$\ eV the lifetime is only
15 fs, a reduction of 60\% from the field-free value.

Recently the phase relaxation time of electrons in the $n=1$ image 
\begin{figure}[t]
\includegraphics[width=67mm,clip=]{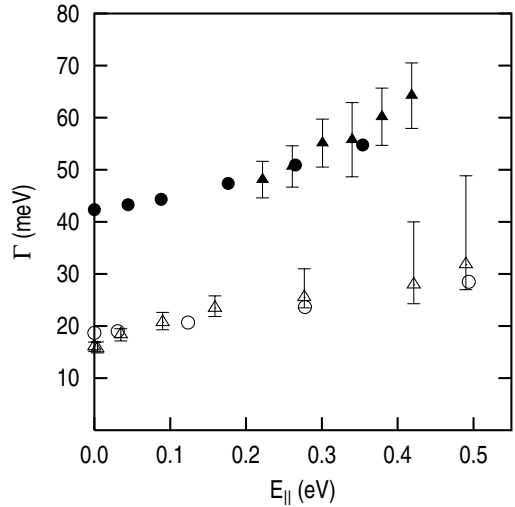}
\caption{
\label{fig:gammak}
Decay rates of electrons in the $n=1$ image state band 
at Cu(100) as a function of lateral energy $E({\bm k}_\|)-E_1$. 
Filled symbols 
correspond to Stark-shifted electrons with energy 
$E_1=4.7$ eV at $k_\|=0$; circles are calculated decay rates and
triangles plus associated error bars are experimental decay rates 
from Ref. [\onlinecite{wah03_}] multipled by 2 to correct for an error in the 
phase relaxation length used in that work \cite{cra05_}.
Open symbols are field-free results; circles are calculated decay rates,
triangles plus associated error bars are decay rates from 2PPE 
experiments \cite{ber02_}.}
\end{figure}
state at Cu(100) has been studied using the STM by Wahl \textit{et al}.
\cite{wah03_},
who measured the spatial decay of quantum interference patterns near
steps. This technique measures the lifetime of electrons with
non-vanishing momentum $\hbar {\bm k}_\|$ 
parallel to the surface, corresponding to
energies above the image state band minimum: 
$E({\bm k}_\|)=E_1+\hbar^2k_\|^2/2m^\ast$.  
In Fig. \ref{fig:gammak} we compare calculated decay rates as a function
of lateral energy with those reported in Ref. \cite{wah03_}. 
There it was concluded that the STM tip did not substantially alter
the dynamical properties of the image-potential states, but 
subsequently an error has been recognised in the identification of the phase
relaxation length that was used \cite{cra05_} so that the values 
displayed in Fig. \ref{fig:gammak}
have been multiplied by 2 to correct for this.
The calculations
are performed for fields which give $E_1=4.7$\ eV. Also shown are 
calculated results
for the field-free case, along with values from 2PPE measurements
\cite{ber02_} which correspond to this case. 
The overall agreement between theoretical and experimental 
lifetimes shown in Fig. \ref{fig:gammak} is very good, and confirms the
existance of a significant field-induced change in the inelastic lifetimes
of the image state electrons.

\begin{figure}[tb]
\includegraphics[width=75mm,clip=]{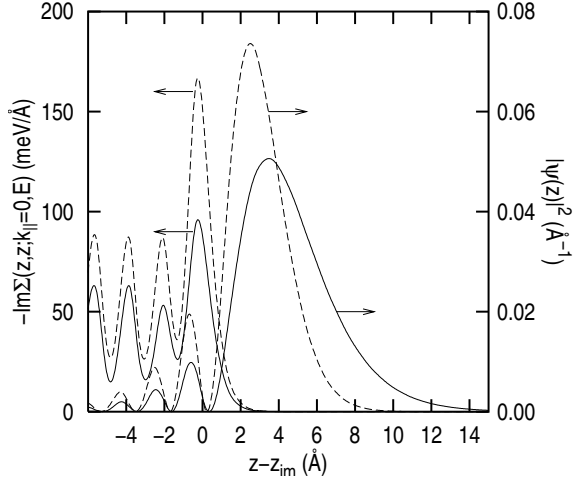}
\caption{
\label{fig:psisig}
Probability density $|\psi|^2$ of the $n=1$ image state
normal to the Cu(100) surface ($\hat{z}$; $z_{\textrm{im}}$ is the position of
the image plane)
and the negative imaginary part of the
self energy $-\textrm{Im}\Sigma$ evaluated for ${\bf k}_\|=0$ and at the
image state energy, shown for $z=z'$.
Solid lines are for the unperturbed image state, $E_1=E_{\textrm{v}}-0.57$\ eV,
and dashed lines for the surface where the image state is Stark-shifted to
an energy $E_1=E_{\textrm{v}}+0.08$\ eV.}
\end{figure}

We now consider the origin of this effect.
In Fig. \ref{fig:psisig} we compare
the probability density $|\psi|^2$ and the imaginary part of the self-energy
$\textrm{Im}\Sigma$ at the Cu(100) surface calculated 
both with and without the 
electric field caused by the tip of the STM.
The changes in $|\psi|^2$ show that
accompanying the Stark-shift to higher energies is a redistribution 
of the weight  of the surface state towards the metal surface, which
increases the spatial overlap with the non-local self-energy.
By displacing the image state electron towards the surface, the
inelastic scattering channels are rendered more efficient.
Calculating $\Gamma$ using the wave function of the Stark-shifted
state but the self energy of the field-free surface at the unperturbed 
image state energy accounts for approximately three-quarters of the full 
increase in $\Gamma$, as shown by the dashed line in Fig. \ref{fig:gamma}.

The remaining change in the decay rate originates in the increase in the
magnitude of $\textrm{Im}\Sigma$ that can also
be seen in Fig. \ref{fig:psisig}. 
We find that calculating the
self-energy using in (\ref{eqn:sigma}) either the 
screened interaction $W({\bm r},{\bm r}';\epsilon)$
of the field-free surface or of the surface in the presence of the 
electric field gives comparable results, i.e., the changes in the electron
wave functions caused by the electric field of the STM tip
do not have a significant effect on the screening of the Coulomb 
interactions that dominate
the inelastic scattering of the image state. Instead, the change
in $\textrm{Im}\Sigma$ is due to the increase in the number of 
final states into which the image state can decay, i.e. the number of states
between $E_F$ and the image state energy, which is Stark-shifted to higher 
energies by the tip-surface interaction of the STM.
Thus the decreased lifetime of the Stark-shifted image-potential 
states results from an increase in the number of final states available 
for inelastic scattering along with increased efficiency of inelastic channels
due to the greater spatial overlap of initial and final state wave functions.

Given the magnitude of the tip-induced change in the lifetimes of
image-potential states it is worthwhile to consider whether 
similar changes affect
STM-derived lifetimes of Shockley surface states 
\cite{jli98_,kli00a,bur99_,bra02_}. Our
calculations for Cu(111) indicate that the effect is minor.
In this case at ${\bm k}_\|=0$ the Shockley state lies at 
$E_F-0.435$ eV which means that
under typical tunneling conditions the electrostatic tip-surface interaction
gives rise to fields that are 5--10 times smaller than 
those present when tunneling into image-potential states,
and the resulting Stark-shift is much smaller: 10-15\ meV \cite{kro04_}.
This causes only a minor change in the number of final states 
that are available for decay.
Furthermore, unlike image potential states which lie predominantly 
outside the surface, much of the Shockley state lies inside the metal, 
and is screened from the tip-induced electric field; there is a negligible
change in the wave function penetration in the presence of the fields.
Overall the electron-electron scattering decay rate which contributes 
two thirds of the total decay rate of $\approx 21$\ meV 
\cite{ech04_} changes by less than 5\% in the electric field.
Although it is safe to dismiss the effect for this particular case, 
it is clear that the tip-induced field will have an
increasingly important effect on the lifetimes of 
Shockley states that lie further from $E_F$, especially
at positive energies.

To conclude, we have used the many-body GW-RPA method to calculate the
inelastic lifetimes of electrons in image states at Cu(100) including the
electric field due to the tip-surface interaction in the STM. We find that 
under typical tunneling conditions the lifetime of electrons in the $n=1$ image
state band is reduced by some 60\% compared to in the absence of the STM tip. 
The Stark-shift to higher energies increases the number of inelastic decay 
channels that are available, whilst the 
the electric field moves the image state electrons closer to the metal 
surface, which significantly increases the efficiency of the scattering 
channels due to increased spatial overlap with final state wave functions.
This tip-induced change in electron lifetimes must be taken into account 
when using the STM to study the dynamical properties of higher-lying 
surface electron states.

\end{document}